\documentclass{sigchi}

\CopyrightYear{2017}
\setcopyright{rightsretained}

\usepackage{balance}       
\usepackage{graphics}      
\usepackage[T1]{fontenc}   
\usepackage{txfonts}
\usepackage{mathptmx}
\usepackage[pdflang={en-US},pdftex]{hyperref}
\usepackage{color}
\usepackage{booktabs}
\usepackage{textcomp}

\usepackage{microtype}        
\usepackage{ccicons}          

\usepackage{todonotes}

\def\plaintitle{A Hierarchical Approach for Investigating Social Features of a City from Mobile Phone Call Detail Records}

\def\emptyauthor{}
\def\plainkeywords{Call Detail Records (CDR); Hierarchical model; Social features.}

\makeatletter
\def\url@leostyle{%
  \@ifundefined{selectfont}{
    \def\UrlFont{\sf}
  }{
    \def\UrlFont{\small\bf\ttfamily}
  }}
\makeatother
\def\pprw{8.5in}
\def\pprh{11in}

\definecolor{linkColor}{RGB}{6,125,233}
\hypersetup{%
  pdftitle={\plaintitle},
  pdfauthor={\emptyauthor},
  pdfkeywords={\plainkeywords},
  pdfdisplaydoctitle=true, 
  bookmarksnumbered,
  pdfstartview={FitH},
  colorlinks,
  citecolor=black,
  filecolor=black,
  linkcolor=black,
  urlcolor=linkColor,
  breaklinks=true,
  hypertexnames=false
}

\begin{document}

\title{\plaintitle}


\numberofauthors{2}
\author{%
  \alignauthor{Fahim Hasan Khan\\
  	\affaddr{Bangladesh University of Engineering and Technology}\\
  	\affaddr{Dhaka, Bangladesh}\\
  	\email{fahimhkhan@gmail.com}}\\
  \alignauthor{Mohammed Eunus Ali\\
  	\affaddr{Bangladesh University of Engineering and Technology}\\
    \affaddr{Dhaka, Bangladesh}\\
    \email{eunus@cse.buet.ac.bd}}\\
}

\maketitle

\begin{abstract}
Cellphone service-providers continuously collect Call Detail Records (CDR) as a usage log containing spatio-temporal traces of phone users. We proposed a multi-layered hierarchical analytical model for large spatio-temporal datasets and applied that for the progressive exploration of social features of a city, e.g., social activities, relationships, and groups, from CDR. This approach utilizes CDR as the preliminary input for the initial layer, and analytical results from consecutive layers are added to the knowledge-base to be used in the subsequent layers to explore more detailed social features. Each subsequent layer uses the results from previous layers, facilitating the discovery of more in-depth social features not predictable in a single-layered approach using only raw CDR. This model starts with exploring aggregated overviews of the social features and gradually focuses on comprehensive details of social relationships and groups, which facilitates a novel approach for investigating CDR datasets for the progressive exploration of social features in a densely-populated city.
\end{abstract}

\category{H.2.8 Database Applications---Data mining, Spatial databases and GIS}\category{I.1.4 Applications}\category{J.4 Social and Behavioral Sciences---Sociology}{}{}


\section{Introduction}

As a consequence of the rapid developments in technologies involving hand-held devices and wireless communications, the usage of mobile phones is increasing day by day. With the rising number of mobile phone users, the cellular networks have become a massive, pervasive sensing system extended over the globe and continuously collecting data. The ubiquitous mobile devices carried by people all over the world are regularly being used as a massive source of spatio-temporal data containing the traces of their daily activities. The modern smartphones have many built-in sensors (GPS, g-sensor, magnetic sensor, etc.) facilitating the collection of spatio-temporal user activity data. But, all the variations of smartphones do not have similar sets of sensors, and activating the sensors for data acquisition is controlled by the users. Moreover, as many people are still using regular mobile phones in developing countries, accumulating data using sensors can be expensive and inconvenient at times. 

Mobile phone service providers continuously collect data in trace files known as Call Detail Records (CDR) from every single active mobile phone of any type using communication services from the towers of their cellular networks. Although CDR data is primarily collected as usage records for billing purposes and network traffic analysis \cite{kalamaras2015mova}, they contain spatio-temporal traces of the mobile phone users. These continuously collected CDR datasets are an existing source of inexpensive data about user activities and being extensively used by many recent research efforts. CDR datasets collected in this process automatically ensure the spontaneous participation of all types of active mobile devices in the network, ranging from the cheapest phone to the most expensive smartphone to function as homogeneous spatio-temporal sensors. As mobile devices are regularly being carried by a huge part of the overall global populace, those are collectively a very effective pervasive sensing platform for acquiring nearly real-time, fine-grained spatio-temporal data. Furthermore, CDR data is collected more effectively with finer granularity in densely populated urban areas and particularly in developed and developing countries, where mobile phone penetration is almost a hundred percent. 

The spatio-temporal responses recorded from mobile device users in CDR are a prospective option for the researchers to formulate various methods for identifying and investigating the city dynamics, vehicular traffic patterns, mobility patterns, and social features of the inhabitants of large cities. Such data is useful for developing applications to facilitate the authority, service providers, and citizens with a better way of understanding, decision-making, discovery, and exploration of urban life. For the last few years, CDR datasets have been decently applied in various researches for understanding different aspects of social activities in the context of urban setups \cite{steenbruggen2015data,hoteit2013estimating,phithakkitnukoon2010activity,lane2010survey}. Analysis of human activity patterns and social behavior is a very important research topic in various fields such as geography, urban and transportation planning, telecom sector, social science, and human psychology. Many of the foundational concepts of sociology and social psychology have originated from the observation of the activities of people living in urban areas. Due to their correlations, examining the daily activities, interactions, and social relationships of people using CDR data leads to the discovery of various social groups enriching the bigger picture of the social features. In social science, a social group is defined as two or more people who interact with one another, connected by social relationships, share similar characteristics, and collectively have a sense of unity. As CDR data can be used to determine the spatio-temporal traces of social interactions, like a person calling another person; and social activities, like the daily travel information of a person in various locations of the city in different times, it is a convenient basis for the analysis of social relationships, interactions and prediction of various social groups to discover the social features of a city.

In this paper, we propose a multilayered hierarchical analytical model to investigate facts from large spatio-temporal datasets and applied that model to build a framework to progressively explore the social features of Dhaka, the highly populated capital of Bangladesh using CDR. The objective of this work is to investigate the CDR data using our multilayered hierarchical exploration model to discover the unique social features of a city, including social activities, interactions, relationships, and social groups. We have applied classifiers and clustering techniques in each of the layers to examine the social features by analysis of the CDR data. In multiple layers, we progressively explore unique social features such as social activities, working and traveling patterns, interactions and relationships, and predict social groups such as regular and irregular working people, extroverts and introverts, frequent travelers, family, friends, co-workers, etc. from the CDR collected from a densely populated urban area. City area features related to social activities like places of common interests, densely populated areas, residential areas, commercial areas, etc. are investigated, and the status of social activities in these areas in different time periods is predicted as well. As the feature of any city area is highly dependent on the underlying social structure and culture, exploration of these features is very much correlated with the investigation of social features.

The novelty of our work is that the proposed multilayered hierarchical exploration model begins data analysis in the first layer with the initial knowledgebase containing the raw CDR data and eventually with the progression of each layer, the knowledge base is augmented by the new information revealed in each of the layers to be further utilized for exploring deeper facts in the subsequent layers. The number of layers can be expanded without any limit, as long as more unexplored information can be predicted from the growing knowledge base. In this work, rather than proposing any new algorithms, we have used a set of existing, optimized, and easy-to-use algorithms based on the shape of the data to be processed in each of the layers in a novel way employing our hierarchical exploration model. Therefore, this approach provides the flexibility of exploration and expandability as the number of layers escalates for learning deeper facts from the CDR data. Although some previous works have investigated various social features using CDR, they lack the flexibility and expandability provided by this multilayered hierarchical model and have a limited scope of exploration because of their non-hierarchical single-layered approaches. To the best of our knowledge, our proposed model and the framework based on that is the only multilayered hierarchical exploration model for analyzing CDR data, which can be further applied on any similar type of spatio-temporal dataset for efficient fact-finding effort. Correspondingly, investigation of the social features of Dhaka, one of the most densely populated cities of a developing country, is a unique study approach for the social analysis of a mega-city.

Finally, we have validated our hierarchical exploration model using a CDR dataset collected from twenty volunteer mobile users with known personal and social information. Another part of validation was done using the information about city area features of Dhaka published by Bangladesh Bureau of Statistics \cite{bbs}.

\begin{figure}
	\centering
	\includegraphics[totalheight=7cm]{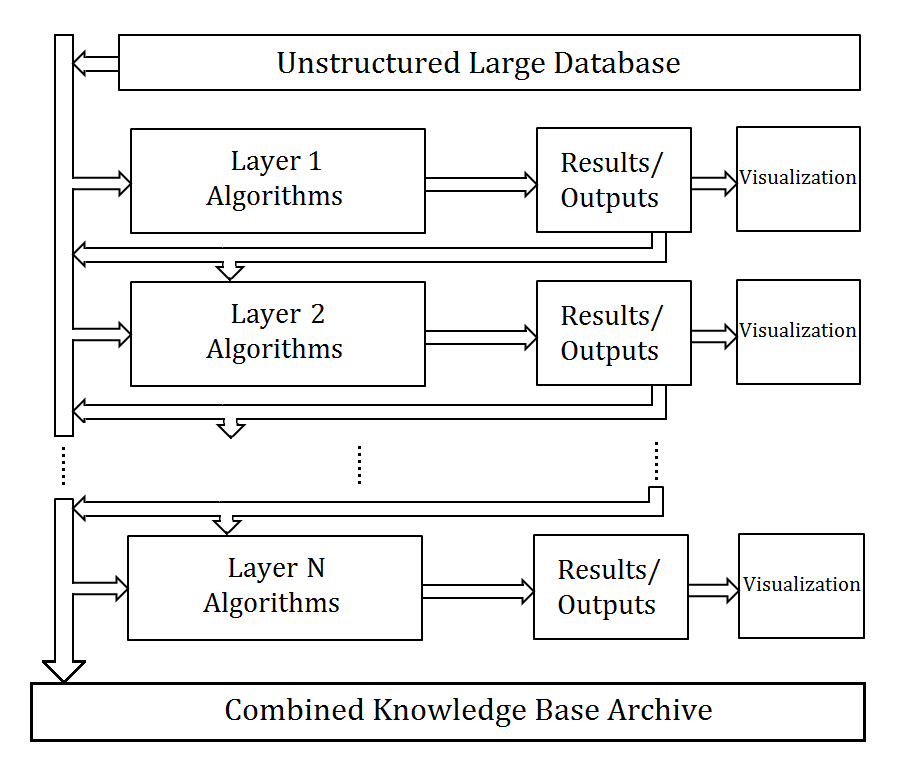}
	\caption{The Multilayered Hierarchical Exploration Model}
	\label{fig:hierarchical}
\end{figure}

\section{Related Works}
Researchers made a good number of approaches for investigating various applications of mobile phone call logs in different forms like CDR. The use of CDR for urban analysis is becoming increasingly popular over the last decade. A survey was conducted on available mobile phone sensing algorithms, their applications, and systems used in many sectors like business, healthcare, social networks, environmental monitoring, transportation, etc. \cite{lane2010survey}. The work of Steenbruggen et al. presented another insightful outline and organized classification of the various studies which utilize mobile phone data for spatial research in a smart city environment \cite{steenbruggen2015data}. As a similar approach, their previous work provided a well-organized outline of researches addressing the use of data derived from mobile phone networks to obtain location and traffic estimations of individuals as a starting point for further research on traffic management \cite{steenbruggen2013mobile}. 

Researchers working on urban and social analysis are taking advantage of using inexpensive CDR for a while. A comparative study between real human trajectory and the one obtained through CDR data in the Boston metropolitan area was conducted using different interpolation methods and taking mobility parameters into consideration\cite{calabrese2011estimating}. A research was done based on a large CDR dataset of approximately one million records of the users in the same region, and a strong correlation was established in daily activity patterns within the group of people who share a common work area's profile \cite{phithakkitnukoon2010activity}. Another work advised that human interaction data obtained using mobile phone Bluetooth sensors can be integrated with human location data obtained through CDR to infer significant human activities information from large and noisy datasets \cite{farrahi2010probabilistic}. A study based on a large dataset from Harbin, China investigated the usage of mobile phones correlates with individual travel behavior \cite{yuan2012correlating}. Similarly, a method was proposed for discovering activity patterns and properties of urban blocks using cell phone data from Beijing \cite{lv2012analyzing}. An analytical study on human activity-travel behavior showed the monthly variability in human activity spaces and locations by analyzing one year's CDR data of Tallinn, Estonia \cite{jarv2014ethnic}. Lately, in another research proposed the utility of spatial correlation patterns for discovering urban and country dynamics from CDR Datasets \cite{trasarti2015discovering}.

Transportation research is another field where CDR was used extensively, and a decent number of researches were done with Origin destination (OD) matrices using mobile phone user data \cite{white2002extracting,white2004use,iqbal2014development,calabrese2011estimating}. The number of vehicles traveling between points on a network over a given period of time can be estimated using an OD matrix. Similarly, many other works were done where road traffic congestion information was detected by applying some methods using mobile phone user data \cite{steenbruggen2013mobile,rose2006mobile}. Caceres et al. published a nice overview of the methodologies of obtaining parameters related to traffic from mobile network-based data\cite{caceres2008review}. A US patent offered a method and gadget for providing vehicular traffic information using existing mobile phone networks \cite{kennedy1996cellular}. An investigative work evaluates the prospects of using mobile phones as traffic probes for obtaining real-time traffic information \cite{rose2006mobile}. Another research found that the variation in the number of mobile phone calls was strongly correlated with the taxi volume of the previous two hours in Lisbon, the capital of Portugal based on data collected from one month of observation \cite{phithakkitnukoon2010taxi}. 

Compared to the existing works, our work proposes the innovative approach of using the multi-layered hierarchical exploration model for investigating a densely populated urban area of a developing country with many distinctive social features. From the literature review of the related works, it is obvious that they focused on investigating a particular social feature using single-layered approaches and no work exist so far which progressively investigate deeper fact by analyzing already discovered facts utilizing a multi-layered hierarchical approach. So, in contrast to other similar works, it is a unique effort to have a fresh look at the social features of the underprivileged society from a completely different angle.

\section{Methodology}
A typical CDR dataset can have hundreds of attributes, containing the spatio-temporal traces of the mobile phone users. Our proposed model functions in multiple layers for processing and analyzing large volumes of CDR data collected from any widespread urban area for progressive discovery of the social features. As this is a generic multilayered hierarchical exploration model, the number of layers is determined based on the available useful attributes in the spatio-temporal database. So, with a higher number of useful attributes, more layers are applicable for exploring the CDR data for finding more facts and keep adding them to the combined database. The combined database can be explored further by applying more layers as long it is possible to obtain new meaningful facts, and it is computationally feasible. So, the general rule for configuring the hierarchical framework for any spatio-temporal database is that the more attributes a dataset have, the more layers can be employed to explore it feasibly for new meaningful facts. 

Each of the layers of our model employs a number of existing, optimized, and easy-to-use algorithms and data analysis methods, including classifiers and clustering techniques designed as single modules to analyze the available information in the knowledge base. In this multilayered hierarchical exploration model, all the modules in every layer function as an autonomous dataset processor for handling the expected input feature vector and producing a district set of output independently. So, new modules can be attached and detached in any layer conveniently for processing new attributes or feature vectors. After finding new facts from each of the layers, they are added to the knowledge base, which also contains the main CDR dataset.

\begin{figure*}
    \centering
    \includegraphics[totalheight=10cm]{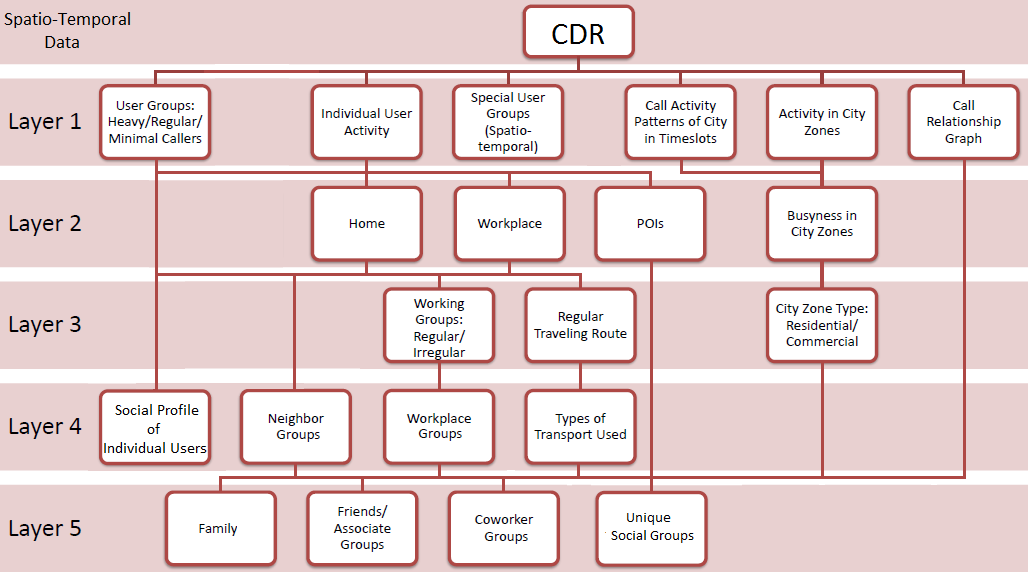}
    \caption{Investigation of Social features of Dhaka City from CDR data in five Layers using the Hierarchical Framework}
    \label{fig:tree}
\end{figure*}

In this work, we explore up to five layers using the available six attributes in our CDR data sets using a framework based on our proposed model(Figure \ref{fig:tree}). The first layer of this framework derives straightforward social features using raw CDR in the knowledge base. The subsequent layer investigates places of interest (POIs), home, and workplace of the users and some features of the different areas of the city. Then the working patterns of the users are examined to identify regular, and irregular working groups, alongside their regular traveling routes in the city, and some more city area features are explored further in the third layer. The following two layers use the facts derived in the initial layers and stored in the knowledge base to progressively identify even closer social groups like family, friends, and acquaintances, neighbors, colleagues, and co-workers, etc. and their social activity patterns and relationships. We keep exploring city area features based on social activities as a parallel but relevant branch of the framework.   

The raw CDR data we used in our work contains six attributes, and they are encrypted user ID; the time, date and duration of the call, and the location (latitude and longitude) of the tower from where the call is made. We have considered each entry of the raw CDR as a 6-dimensional feature vector $x_{i}$ where $i = 1,2,...,n$ and $n$ is a positive integer indicating the number of entries in the available CDR. Each The features contained in each  $x_{i}$ are $(u, dt, tm, dur, loc(lat, lon))$. The set of feature vectors containing the full CDR data is $X$, where

\begin{equation}
    X = \{x_i|\{u, dt, tm, dur, loc(lat, lon)\} \in x_i, i \in Z+\}
\end{equation}

Where the notations are,\\ \\
$u$ = Encrypted User ID assigned to every user\\
$dt$ = Date of a call/communication activity\\
$tm$ = Time of starting a call/communication activity\\
$dur$ = Duration of a call\\
$lat$ = Latitude of the cell tower providing the call \\
$lon$ = Longitude of the cell tower providing the call \\
$loc$ = A location consisting of a (lat, lon) pair\\

\subsubsection{Layer 1}
In this layer, we predict some straightforward social features using statistical analysis and simple linear classifiers. At first, CDR logs for every unique user is generated summarizing his call activities in all distinct visited locations for the full-time period available in the CDR data. These individual user logs are stored in the knowledge base to be used in the next layers. Then, we determine the daily intensity of user activities and the calling pattern of users over different time periods. On the basis of the concentration of monthly call activities, we classify three groups named as minimal users, regular users, and heavy users using a linear classification. First, we calculate the usage score, $\mu$, of every user using the following equation,

\begin{equation}
    \label{eq:usg3}
    \mu_u=\omega_c NC_{T,loc}+\omega_d \sum dur_{T,loc}
\end{equation}

Where, 

$T$ is the time period of call activity for user $u$.

$loc$ is the location of call activity for user $u$.

$NC$ = number of calls made by the user in time period $T$

$\sum dur$ = total duration (in seconds) of calls made by the users in time period $T$\\

$\omega_c$ and $\omega_d$ are constant weight factors.

We include every user in any of the three groups based on his $\mu$. This classification shows us the overview of the mobile phone usage pattern of a city and characterizes an important social feature signifying social interactions and activeness of people. Investigation of special user groups with known facts based on-call activities is done by calculating $\mu$ for different time periods and locations. For example, a group of people talks noticeably more in business hours. We can deduce that these people are a group of professionals whose business requires a lot of communication with people. 

As all user ID is encrypted for anonymity and the call destination information is not present in our CDR, a call relationship graph is generated, which is critical for the analysis in the next layers. We have the time and duration of every call activity of the user. By cross-referencing the whole database, we detect the call destination for every entry and generate the calling relationship graph among the users in our CDR. The number of calls made by two users is assigned as the weight of the calling-relationship edge of two user nodes.

\subsubsection{Layer 2}
The locations of home, workplace, and other frequently visited places for every single user is predicted and stored in the knowledge base archive in this layer. In this text, we mentioned home, workplace, and other frequently visited places commonly as "Place of Interest" or POI. Each user call log created in the first layer is used to determine the top POI locations from where he makes most of the calls activity determined by his usage score $\mu$ for those locations. Then, from the maximum usage score, $\mu$ in usual working hours and off-hours leads to the discovery of the home and workplace of a user, respectively, from the list of POIs. Other POIs are locations where the user visits frequently. It is safe to assume that all user has a home location, a good number of users, who are working people, have a workplace location, and many users have one or more regularly visited POIs. We have used clustering techniques to find the POIs of a user. Every sizable cluster from the list of clusters detected by a clustering algorithm is considered as a POI. From the various clustering algorithms, we have selected X-Means and EM clustering algorithms to find the POIs based on the performance of our experimental result. Usually, people spend their off-hours or the night time, in their home. From this knowledge, we predict the home of a user by comparing his usage score $\mu$ at every POIs at off-hours in the total time period of the available CDR data. Here, the usage score is calculated using the following equation, 

\begin{equation}
    \label{eq:usg4}
    \mu_{poi}=\omega_c NC_{T,poi}+\omega_d \sum dur_{T,poi}
\end{equation}

Where,

$NC$ = number of calls made by the user in $T$ period from a POI

$\sum dur$ = total duration (in seconds) of calls made by the users in T period from a POI

$T$ = time period of working hours/ off hours of the city\\

$\omega_c$ and $\omega_d$ are constant weight factors, whose values are tuned up using a linear classifier.

For every city, the usual working hours and off-hours is a known fact. Obviously, the POI with the highest usage score in off-hours is predicted as the home of any user. Similarly, workplaces are predicted by finding usage score of POIs in working hours to predict workplaces of the users. The important thing to be considered while predicting the workplace is that everyone may not have a well-defined workplace. For some of them, the maximum valued POI detected during WORKING-HOURS also indicates home and easily identified by matching the location of maximum valued POI detected during WORKING-HOURS with the location of the home. Besides these two prominent social groups, there are other types of people with complex home and workplace pattern. 

Furthermore, city areas are classified as BUSY or IDLE based on their level of activity in different time periods using a linear Support Vector Machine (SVM). The CDR has a fixed number of unique locations, and we consider them as different zones of the city. The concentration of call activities and the number of active users at different times of the day determines how busy or densely populated a zone is at a certain time period of weekdays and weekends.

\subsubsection{Layer 3}

Using the home and workplace information from the previous layer, at first, we distinguish two major social groups based on the working activity patterns; the working people, who regularly goes to a certain workplace and irregular working people, who either stays home or have some indistinguishable irregular working pattern. The group of regular working people spends their working-hours in their office and off-hours in their home. People from this group regularly travel from their home to the workplace and have a recognizable traveling pattern. People like professionals, office workers, businessmen, and students belong to this group. The irregular working group includes home staying people like homemakers, retired, and unemployed people. 

We have used a linear classifier based on SVM to predict these two major social groups. In this layer, classification is done based on the traveling distance of the regular working people from home to workplace. We have used the Haversine formula to calculate the great-circle (surface of Earth) distances between the home and workplace from their longitudes and latitudes, where the coordinates of the two locations are $(lat1,lon1)$ and $(lat2,lon2)$. We predict the regular traveling route of a user by considering all his unique call locations during his travels between home to the workplace. It is done by applying Dijkstra's algorithm for finding single-source shortest paths considering the unique call locations as nodes, home location as the source, and workplace as the destination. To do so first a graph $G(V,E)$ is created, where,

$V\leftarrow {loc_{1},loc_{2},...,loc_{i}}$, List of unique call locations of the user en route home to workplace.

$E\leftarrow$ edges representing all possible paths between ever pairs of nodes created based on real-world map data and the distance between them are the values of edges.

We also predicted the regular traveling route of a user by considering his call locations during his travels between home to the workplace using the shortest-path finding algorithm. 

Additionally, we also classify the city zones as Residential, Commercial, and Miscellaneous as social activities and interactions are considerably different in each type of area. We have used a linear SVM to classify the city zones.

\subsubsection{Layer 4}
In the fourth layer, the knowledge base is already augmented with a good amount of derived facts from the earlier layers, and we start investigating more significant social features by narrowing down the social groups and relationships based on some universal and well established social facts. 

It is a known fact that a group of users having the same home locations means that they live in the same neighborhood. So, we classify these loosely connected social groups as neighbors; and members of the same group have a high probability of interaction. Using this fact, we predict the neighbor groups in the city. Similarly, if a group of users has the same workplace locations, they have a high probability of being acquaintances, even colleagues. Another well-known fact is that people working in the same area may know each other; they may even work in the same office, making them a group of office workers. For example, the people who have businesses or shops in the same area usually have a kind of mutual relationship, which indicates a social group of businessmen.

The type of transport used by a user can be predicted and classified from his traveling route and information on time differences and distances among call activities made by him on the same day on that route. From this fact, we have predicted the type (based on speed) of transport used by users in their trips through traveling routes. To do so, we calculate the speed of transport by calculating the distance between two locations and the time difference from consecutive CDR entries. Then, we calculate the average speed and use these features in an SVM to predict the type of transport.

If two or more persons have the same home, workplace, and some common POIs and overlapping regular traveling routes, we can perform predictive analysis of their social relationship and group membership using this available information. We have applied this fact to predict social relationships based on the probability of interactions, work and travel patterns. For example, if a group of people shares the same home and workplace, we can infer that they are co-workers and live in the same residential facility, which may be provided by the employer. This inference would be more established if we find a similar traveling pattern, which we detect by predicting this traveling route and time period of traveling. 

In this layer, we create an individual user's social profile based on the collective information up to this layer. By changing the parameters for location and time period in Equation \ref{eq:usg3} based on the known fact of a city, we examine the social activities and find different group membership of a user. For example, the users with high usage score in the time period from 12 a.m. to 4 p.m. are late-night callers. Similarly, working user with high usage score during working hours belongs to a special group of professionals who highly focus on communication. As an example of location-based prediction, if the users have the workplace in a university area, they have a high probability of being a student or teacher of that university. Similarly, a user with the workplace in a large shopping mall is likely a shopkeeper, a user with the workplace in an army base is a prospective member of the army, and the user with a workplace in a hospital area is probably a doctor, nurse, or medical personnel. Furthermore, we predict the working days and off days of a regular working user from his days spent at the workplace during working hours.

\subsubsection{Layer 5}
In the fifth layer, we use the calling relationship graph to narrow further down social relationships and groups, which includes family, friends, colleagues, and closely acquainted people. A calling-relationship is the strongest indication of any social relationship. Also, call duration, number, and frequency of calls made are proportional to the closeness of the social relationship. People with close social relationships call each other more often and often talk for a longer time. For calculating the calling relationship, we considered the time of call, call duration, number, and frequency of calls made. For instance, if two or more people share the same home location and have a frequent calling relationship, we can predict the probability of them being family members or close acquaintances. The closeness of this type of relationship was further investigated by examining the time and duration of calls and the frequency of calls made. Similarly, all the people who work in the same area can be considered as colleagues or friends based on their calling relationship patterns. If two users from the same workplace call each other frequently in the off-hours, it indicates a closer social relationship. 

We can further generalize the fact that if two or more persons visit or stay in the same POIs multiple times in the same time periods and have a strong calling relationship, we predict them as being a member of a social group like family, friends or other types of close relationships. On the aggregated knowledge base containing predicted information of all the users, we apply the chain rule of probability to predict the social closeness among users and discover different social groups based on social closeness.

\section {Results and Analysis}

\subsection{Dataset}

CDR datasets can have hundreds of attributes based on the telecom equipment and configurations used by the cell operators. This research is done using CDR datasets of the mobile phone users from Dhaka, the capital city of Bangladesh and collected by Grameenphone Ltd, a major telecom operator of Bangladesh. The dataset comprises 971.33 million anonymous CDR entries obtained from 6.9 million users, which is more than 55 percent of the total population of the study area at that time. The total duration of the dataset is one month, starting from June 19, 2012 to July 18, 2012 collected using the cell towers from 1360 unique tower locations distributed all over Dhaka city. Each record of our CDR Dataset has the following attributes: a random ID number of the phone; the exact time and date; call duration and location (latitude and longitude) of the cell tower that provided the network signal for the mobile device activities. Although this CDR dataset has very limited number of attributes, they are some commonly used key attributes in all variations of CDR data. Dhaka is a densely populated city with frequent communication activity and has greater number of the cell towers providing good coverage resulting CDR data of good granularity. The major limitation of CDR dataset is the spatial precision which is dependent on the number and location of the tower. The total land area of Dhaka city is 280 sq-km, which gives us an average of 5 tower locations per sq-km which is good for providing a better than sq-km scale precise location for each users.

\subsection{Results}
The framework based on the hierarchical exploration model is implemented using Java to process the CDR data, compute the results and some basic visualization. In this section, the results are presented and discussed as they are obtained in different layers of our hierarchical exploration model.

\subsubsection{Layer 1}
\begin{figure}[h]
	\centering
	\includegraphics[totalheight=6cm]{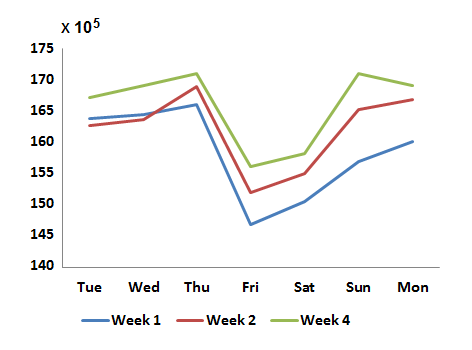}
	\caption{Comparing call traffic variations from usage score, $\mu$ in three consecutive weeks. From this figure it is correctly inferred that Friday is the weekend of Dhaka City.}
	\label{fig:wkver}
\end{figure}

In the first layer, analysis on raw CDR data to detect the call activities of all users in different time period of the day gives us insight about the intensity of activities in the city during different periods of the day, as phone calls are directly related to many social activities. For example, in our CDR data of one month, maximum number of active users and calls is found on 5th July, 2012, which was a national holiday in Dhaka city. So, it is inferred that people make social interaction over phone more on holidays. On the other hand, it is also observed that people make less call on weekends (Figure \ref{fig:wkver}). These are two significant social features of a city. Also, we have detected the call activities in different time periods of the day, which reflect the overall activities and city dynamics of the given time periods as shown in (Table \ref{table_4}). We divided the days in four time slots and identified the user activity accordingly.

\begin{table}[h]
	\caption{User activities in different time periods of the day}
	\label{table_4}
	\centering
	\begin{tabular}{ | l | c | c | }
		\hline
		Time period & No of Calls & No of Active Users \\ \hline
		Night (12 to 6 AM) & 370311 & 192928 \\ \hline
		Evening (6 to 10 AM) & 1256755 & 578975 \\ \hline
		Midday (10 AM to 5 PM) & 3442933 & 960704 \\ \hline
		Morning (5 PM to 12 AM) & 3187238 & 968891 \\
		\hline
	\end{tabular}
\end{table}

The call volume shows strong variations with time and day of the week, (Table \ref{table_4}), but differences across subsequent weeks are generally mild considering the call traffic of the whole city (Figure \ref{fig:wkver}).

From the usage score, $US$, we classify three user groups, Minimal users, Regular users and Heavy users. From this classification we can inferred that, the heavy users are socially more active or extroverts, while other two groups are less active or introverts proportionately. This fact is useful for investigating the correlation between mobile phone usage and social activeness.

\begin{table}[h]
	\centering
	\caption{Classifying users based on call activity}
	\label{call_activity}
	\begin{tabular}{|l|c|c|}
		\hline
		\textbf{Group} & \textbf{No of Users} & \textbf{Percentage} \\ \hline
		Minimal users   & 2478480              & 36                  \\ \hline
		Regular users   & 4036447              & 58                  \\ \hline
		Heavy users     & 412046               & 6                   \\ \hline
	\end{tabular}
\end{table}

\subsubsection{Layer 2}
\begin{figure*}
	\centering
	\includegraphics[totalheight=6cm]{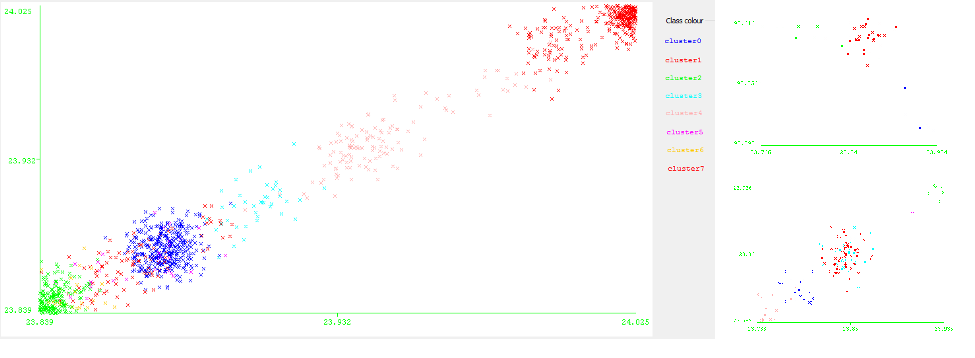}
	\caption{Prediction of POIs, Home and Workplace using EM Clustering Algorithm}
	\label{fig:POI_EM}
\end{figure*}

The knowledge base augmented with facts from first layer is used for finding the POIs of the users in this layer. Figure \ref{fig:POI_EM} is showing the clustering approach for predicting POIs, home, and workplace of a user using EM clustering algorithm.

\begin{figure}[!h]
	\centering
	\includegraphics[totalheight=12.5cm]{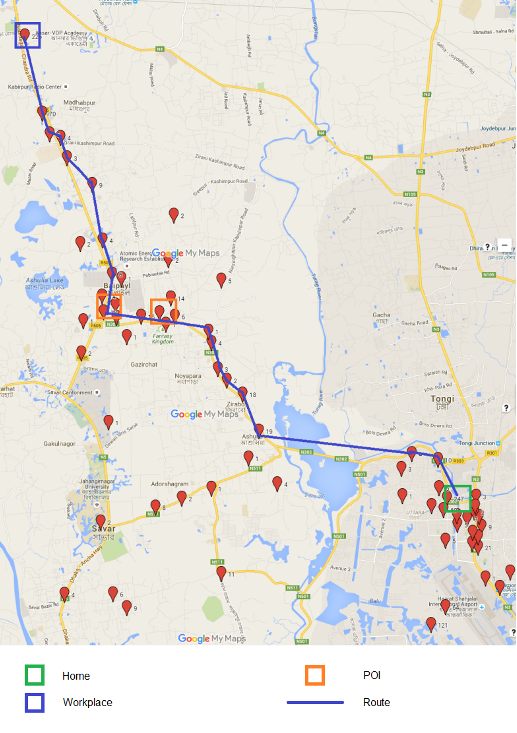}
	\caption{Home, Workplace, POIs and Regular Traveling Route of an user}
	\label{fig:route}
\end{figure}

We predicted the home and workplace of the users from the POIs and the known usual working hour of Dhaka city, which is 9 AM to 5 PM or a slight variation of this time-slot. On the map of Figure \ref{fig:route}, we can see the visualization of the home, workplace and POIs of a user. Map markers are placed in the locations from where he has made one or more calls and beside every marker the number of calls made from that location is mentioned.  

\subsubsection{Layer 3}
In this layer we process the knowledge base archive to find the home and workplace information of all the users using our clustering algorithm iteratively, resulting another large database to be added to the knowledge base. A sample of predicted results for some users is shown in Table \ref{HWtable}.

\begin{table*}
	\centering
	\caption{Predicted information of a few users form the CDR data}
	\label{HWtable}
	\begin{tabular}{|c|c|c|c|c|c|c|c|c|}
		\hline
		User ID   & \multicolumn{2}{c|}{Home} & \multicolumn{2}{c|}{Workplace} & Distance & NCall & Dur Call & US   \\ \hline
		AAH86JAa9 & 23.789      & 90.408      & 23.787         & 90.415        & 0.72     & 22    & 3469     & 80   \\ \hline
		AAH86JAa7 & 23.796      & 90.364      & NA             & NA            & NA       & 966   & 78429    & 2273 \\ \hline
		AAH86JAA8 & 23.707      & 90.410      & NA             & NA            & NA       & 165   & 18731    & 477  \\ \hline
		AAH86JAA9 & 23.846      & 90.421      & 23.793         & 90.402        & 6.20     & 50    & 6262     & 154  \\ \hline
		AAH86JAAA & 23.710      & 90.404      & 23.812         & 90.255        & 18.93    & 26    & 4201     & 96   \\ \hline
		AAH86JAa8 & 23.723      & 90.384      & NA             & NA            & NA       & 101   & 23433    & 492  \\ \hline
	\end{tabular}
\end{table*}

From the SVM based classifier applied on home and workplace database of all users, we found two groups of people based on their working patterns. One group, the regular workers, has a consistent traveling patterns between their homes and workplaces. The rest of the people has irregular mobility patterns and no specific workplaces. So, we considered them as irregular workers. Subsequently, when the home and workplace of the regular working user group is found, we have used this information to calculate the distances regularly traveled between their homes and workplaces.

\begin{table}[h]
	\renewcommand{\arraystretch}{1.5}
	\caption{Working pattern of the Users: Regular vs Irregular}
	\label{table_3}
	\centering
	\begin{tabular}{ | c | c | c | }
		\hline
		User with irregular working pattern & 5163239 & 74.5 Percent \\ \hline
		User with regular working pattern	& 1763734 & 25.5 Percent \\
		\hline
	\end{tabular}
\end{table}

From the results, we observe that 1.8 million (25.5 percent) of the 6.9 million users of our CDR data have consistent working schedules and the homes and workplaces of these people is clearly detected. The other 74.5 percent people have irregular patterns of home and workplace as shown in Table \ref{table_3}. From our regional knowledge of Dhaka city, we can deduce that, people like housewives, retired people, part-time and irregular workers, etc. belong to this group. For all user with regular working pattern, traveled distances for attending workplace from home is calculated and we classified them in five subgroups for understanding their mobility and traveling pattern.

\begin{table}[h]
	\caption{Traveling distance to workplace for regular workers}
	\label{table_5}
	\centering
	
	\begin{tabular}{ | c | c | c | }
		\hline
		Traveling Distance & No of User & Percentage\\ \hline
		0-2 km & 951217 & 53.93 \\ \hline
		2-5 km & 368484 & 20.89 \\ \hline
		5-10 km & 258620 & 14.66 \\ \hline
		10-20 km & 149168 & 8.46 \\ \hline
		20-100 km & 36245 & 2.06 \\
		\hline
	\end{tabular}
\end{table}

From the findings in Table \ref{table_5} it is evident that, 53.93 percent of the regularly working people live within two kilometers of their workplace and only 2.06 percent people live more than 20 km away from their workplace. We can infer a social feature that, most of the people in Dhaka try to stay close to their workplaces and travel less for going to the workplaces. Consequently, they usually select their home near their workplace, as traveling in a densely populated city like Dhaka is difficult. 

We have predicted the regular traveling route of a user by considering his calling locations during his travels between home to workplace.  The traveling route of a random user is shown in Table \ref{troute} which can be visualized in a map as presented in Figure \ref{fig:route}.

\begin{table}[!htbp]
	\centering
	\caption{Traveling Route of a user}
	\label{troute}
	\begin{tabular}{|l|l|l|}
		\hline
		LAT       & LONG      & Remarks            \\ \hline
		23.878599 & 90.390602 & Home               \\ \hline
		23.843901 & 90.279404 &  					\\ \cline{1-2}
		23.848301 & 90.274696 &                    \\ \cline{1-2}
		23.864401 & 90.3992   &                    \\ \cline{1-2}
		23.875799 & 90.289398 &                    \\ \cline{1-2}
		23.8792   & 90.400597 &                    \\ \cline{1-2}
		23.883101 & 90.331703 & Traveling Route                   \\ \cline{1-2}
		23.8908   & 90.387497 &                    \\ \cline{1-2}
		23.937799 & 90.2714   &                    \\ \cline{1-2}
		23.9606   & 90.271103 &                    \\ \cline{1-2}
		23.9781   & 90.267502 &                    \\ \cline{1-2}
		23.9928   & 90.256699 &                    \\ \hline
		24.025    & 90.244202 & Workplace          \\ \hline
	\end{tabular}
\end{table}

\subsubsection{Layer 4}
In the fourth layer, using the home information of the users we detect the group of people who live in same neighborhood and from the workplace location we predict the working area based group who has the likelihood of having professional relationship. We build up social profile of users based on the facts detected in the previous layer about of home, workplace, POIs, regular traveling routes and working pattern. Two examples of predicted user profiles are shown in Table \ref{user profile}.

\begin{table}[!htbp]
	\centering
	\caption{Examples of two predicted user profile}
	\label{user profile}
	\begin{tabular}{|l|c|c|c|}
		\hline
		User ID (Encrypted)     & P4EAcw               & BBDAYO             \\ \hline
		User type               & Regular Worker       & Irregular Worker   \\ \hline
		Home                    & 23.7035, 90.4563 & 23.8106, 90.3714 		\\ \hline
		Workplace               & 23.9508,90.2714      & NA                 \\ \hline
		Traveling distance      & 33.31                & 0                  \\ \hline
		Working Hours 			& 10 AM to 6 PM        & NA                 \\ \hline
		Off days      			& Friday               & NA                 \\ \hline
		Social Group  			& Service Holder       & Homemaker          \\ \hline
	\end{tabular}
\end{table}

\subsubsection{Layer 5}
\begin{figure}[h]
	\centering
	\includegraphics[totalheight=7cm]{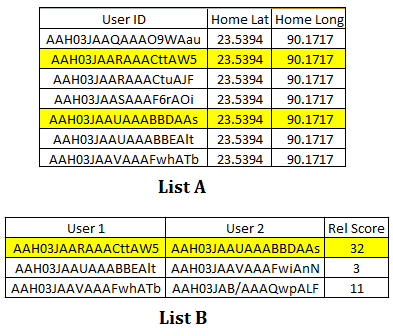}
	\caption{Finding family members from home location and calling relation}
	\label{L52}
\end{figure}

Using the calling relationship information on the aggregated knowledge base we predict family, friends, colleagues and closely acquainted people. For example, if two or more people share the same home location and have a frequent calling relationship, we deduce them as being family members or close acquaintances. Figure \ref{L52} illustrates an instance of such relationship. From List A and List B we can see that the two highlighted users have same home location and they have a strong calling relationship. Therefore, it is strongly inferred that, the highlighted users are closely related and belongs to the same social group.

\section{Validation}
By applying our hierarchical exploration model on the large CDR dataset, we have discovered quite a few facts on the social features of Dhaka city. For validating our model for predicting social facts, two different sources have been used. The two sources of validation are a personal CDR dataset collected from twenty volunteer mobile phone users and a report \cite{bbs} published by Bangladesh Bureau of Statistics, the detail of which is discussed in this section.

\begin{table}[h]
	\centering
	\caption{Accuracy of Prediction}
	\label{validation}
	\begin{tabular}{|l|c|}
		\hline
		\textbf{Predicted Facts} & \textbf{Success Rate (Percent)} \\ \hline
		Home Location           & 100                         \\ \hline
		Workplace Location      & 100                         \\ \hline
		Working Group           & 80                          \\ \hline
		Regular Traveling Route & 75                          \\ \hline
		Caller Type             & 90                          \\ \hline
		Social Groups (Any)      & 100                         \\ \hline
	\end{tabular}
\end{table} 

Some of the social facts (as in Table \ref{validation}) derived from our hierarchical exploration model are related to the personal information of the user, which is not possible to validate in our main CDR dataset due to the fact the user ids are encrypted by the cell phone operator. For validating effectiveness of our model for these social facts, we collected the personal CDR data of one month from twenty randomly selected volunteer participants with known personal information, who are using mobile phone service from the same cell phone network operator who provided us the large CDR database. This operator provides the users with the option to download CDR data of their own cell phone usage by logging into a personal service information website. The volunteer cell phone users provided us their personal CDR data of one month, which is used for creating a small validation dataset and we applied our hierarchical data exploration model on the validation dataset to find the same social facts (as in Table \ref{validation}) we explored the large encrypted CDR dataset in five layers. The accuracy of our hierarchical exploration model for finding fact from the validation dataset is summarized in Table \ref{validation}. For calculating the success rate of social groups detection, if our model identified the user as member of at least one of the social groups from Figure \ref{fig:tree}, we considered it as a successful prediction.

\begin{figure*}
	\centering
	\includegraphics[totalheight=5cm]{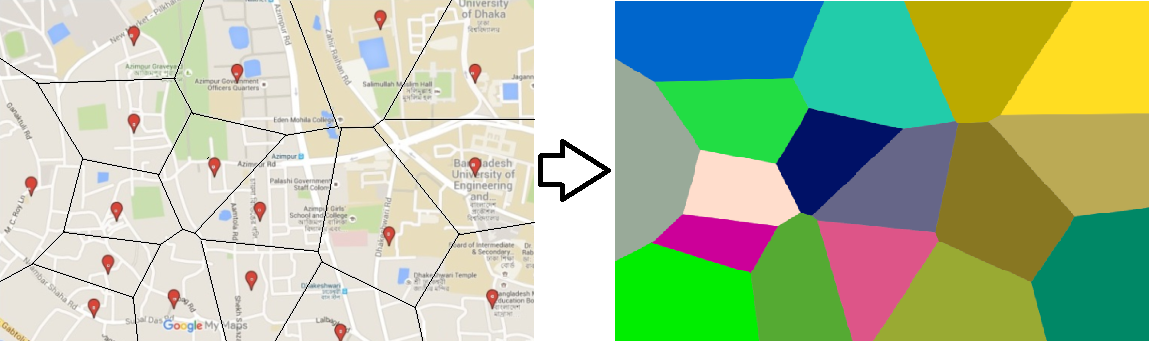}
	\caption{Example of visualizing attributes of different areas in a part of Dhaka City using Voronoi Maps}
	\label{fig:vis}
\end{figure*}

\begin{table}[h]
	\centering
	\caption{Example of predicting city area type}
	\label{predarea}
	\begin{tabular}{|l|l|l|}
		\hline
		\textbf{Location}    & \textbf{Prediction} & \textbf{Validated from Map}    \\ \hline
		23.750,90.359  & Residential        & Kaderabad Housing             \\ \hline
		23.729,90.384  & Residential        & Azimpur Govt. Offr. Quarter \\ \hline
		23.707, 90.439   & Commercial         & Jatrabari Bazar               \\ \hline
		23.740,90.373  & Residential        & Dhanmondi Residential Area    \\ \hline
		23.755, 90.389 & Commercial         & Farmgate Intersection         \\ \hline
		23.833,90.4153    & Residential        & Nikunja-2 Residential Area    \\ \hline
		23.871,90.390    & Residential        & Uttara Residential Area       \\ \hline
		23.781, 90.406     & Commercial         & Mohakhali Commercial Area     \\ \hline
	\end{tabular}
\end{table}

On the other hand, we explored the social features of the different areas of the city and determined their types. To validate our model for detecting the social features of the different areas of the city, we grouped the available 1360 unique location from the large CDR dataset based on the administrative area map of Dhaka city provided by the report, "District Statistics 2011 Dhaka" \cite{bbs}. Dhaka city has a total of 119 wards (administrative area), which we mapped with the locations from the main CDR dataset. We visualized the busyness of various areas of the city in different time of the day represented by Voronoi diagrams generated on the map (Figure \ref{fig:vis}). The city zone type is also validated with the map and data presented in the report \cite{bbs}. Among the 1360 locations, we correctly predicted city area types of 1257 locations, which makes the accuracy 92 percent. Some example of the predicted city zone and their validation is presented in Table \ref{predarea}.

\section{Conclusion and Future Works}
In this research work, we introduced a novel multilayered hierarchical model for analyzing large spatio-temporal dataset and applied that model for progressive exploration of social activities, relationships and groups from CDR and investigated the unique social features of Dhaka, one of the most populated city of the world. This proposed generic framework utilizes a set of existing, effective and easy to use algorithms for analyzing any spatio-temporal or high dimensional large data set using multiple layers for fact finding. Here, the model is used for exploring social relationships and groups starting from an aggregated level to more detailed social features progressively. Using this model, we predicted social features which includes, social groups based on home, workplace, and places of interest, working and traveling pattern, and relationships among family, friends, colleagues and acquaintances from CDR data. Also, we explored city area features such as places of common interests, residential and commercial areas, and activity patterns in different city areas. Our effort offers to overcome the limitations of previous works by providing a new model with the expandability and flexibility of progressively exploring larger datasets and a fresh look towards the social features of a highly populated city of a developing country.

As a future work, we want to experiment with CDR of longer durations with wider ranges of attributes from Dhaka and other cities as well to investigate and compare their unique social features using our multilayered hierarchical model. Also, this model can be applied as a universal model on other types of large spatio-temporal datasets to explore other potential applications. In addition, we have implemented our model using the traditional methods of data analysis. As the simultaneous processing of CDR data from multiple users performed in different layers are highly parallel, the emerging technology of using general-purpose Graphics Processing Unit (GPGPU) for parallel processing can provide us a huge performance boost up. Therefore, we have plan for implementing our model using GPGPU support for faster and optimized performance.

\balance{}

\bibliographystyle{SIGCHI-Reference-Format}
\bibliography{hierarchical}

\end{document}